\documentclass[aps,prl,superscriptaddress,twocolumn,floatfix,showpacs]{revtex4}
\usepackage{graphicx}
\usepackage[]{amsmath}

\begin{document}

% number in the upper right hand corner of the title page in preprint mode.
%\preprint{First Draft}

\title{Universal scaling relations in molecular superconductors}

\author{F. L. Pratt}
\email{f.pratt@isis.rl.ac.uk}
\affiliation{ISIS Facility, Rutherford Appleton Laboratory, 
Chilton, Oxfordshire OX11
0QX, United Kingdom}
\author{S. J. Blundell}
\email{s.blundell@physics.ox.ac.uk}
\affiliation{Clarendon Laboratory, University of Oxford, Parks Road,
  Oxford OX1 3PU, United Kingdom}
%\thanks{}
%\affiliation{}


\date{\today}

\newcommand{\chem}[1]{\ensuremath{\mathrm{#1}}}

\begin{abstract}
Scaling relations between the
superconducting transition temperature $T_{\rm c}$, the  
superfluid  stiffness $\rho_{\rm s}$ and the normal state conductivity $\sigma_0(T_{\rm c})$
are identified within the class of molecular  superconductors.
These new scaling  properties
hold as $T_{\rm c}$ varies over  two orders  
of  magnitude  for  materials  with
differing  dimensionality  and  contrasting  molecular  structure,  and  are
dramatically different from the equivalent scaling properties observed within the family of cuprate superconductors. 
These scaling relations place strong constraints on theories for molecular superconductivity. 
\end{abstract}

\pacs{
76.75.+i, %% musr
74.25.Nf, %% sc response to em fields
74.25.Fy, %% sc transport properties
74.70.Kn, %% organic sc
74.20.De  %% pheonomenological sc theories
 } 
%\keywords{}

\maketitle


Understanding the phenomenon of superconductivity,  now  observed  in  quite
disparate  systems,  such  as  metallic  elements,  cuprates  and  molecular
metals, involves searching for universal trends across different  materials,
which  might  provide  pointers  towards  the  underlying   mechanisms.    
One such trend is the linear scaling between the
superconducting transition temperature ($T_{\rm c}$)  and  the  
superfluid  stiffness ($\rho_{\rm s} = c^2/\lambda^2$, where $\lambda$ 
is the London penetration depth), first 
identified by Uemura {\sl et al} for the underdoped cuprates \cite{uemura}.   
Recently,  scaling
relations between $\rho_{\rm s}$ and the normal state conductivity 
$\sigma_0$ have also been suggested and
a linear relation between $\rho_{\rm s}$ and the product 
$\sigma_0(T_{\rm c})T_{\rm c}$ was  demonstrated  for
cuprates and some elemental superconductors \cite{homes}.
Here we show that these specific linear scaling relations do not hold for molecular superconductors,   
but a different form of power-law scaling is found to link $\rho_{\rm s}$, $\sigma_0(T_{\rm c})$ and  $T_{\rm c}$.  
These  scaling  properties hold as $T_{\rm c}$ varies over  several  orders  of  magnitude  for  
materials  with differing  dimensionality  and  contrasting  molecular  structure,  and  the scaling is
dramatically different  from  that  of  the  cuprates.   Our  findings  have
considerable implications for the theory of superconductivity in molecular 
systems.

Molecular superconductors are generally regarded as  members  of  the  wider
group of `exotic' superconductors that have attracted much  research  effort in recent years. 
However, the  number  of  different examples  of 
molecular superconductors is now sufficiently large that  systematic  studies  of  their
properties may be made  independently  of  the  other  non-molecular  exotic superconductors.  
A general feature of all these exotic  superconductors  is
the large carrier scattering rate observed in the normal state \cite{zaanen}  leading  to
a picture of them as `bad metals' \cite{emery}.
The scattering  rate  at  temperatures near $T_{\rm c}$ may have particular relevance for the  superconductivity,  
since  it is expected that similar carrier interaction mechanisms  would  be  dominant
in the normal state resistance and in the pairing of carriers that leads  to
the formation of the superconducting  state.   
It  is  therefore  useful to study the correlation between $\sigma_0(T_{\rm c})$ 
and superconducting  parameters such as $T_{\rm c}$ and $\rho_{\rm s}$.  
Fig.~1(a) shows $\rho_{\rm s}/c^2$ ($=ne^2/m^*\epsilon_0c^2$)  and  $T_{\rm c}$  derived
from  $\mu$SR  measurements in the vortex state \cite{sjb,sonier},  
plotted   against   $\sigma_0(T_{\rm c})$   in   the   highest
conductivity direction for  a  series  of  molecular  superconductors.   
The materials  range  from  a  highly  anisotropic  quasi-one-dimensional  (q1D)
organic superconductor ((TMTSF)$_2$ClO$_4$), through  systems of two-dimensional
(2D) layered organic superconductors (BETS and  ET  salts) to examples  of
three-dimensional (3D) fulleride superconductors; full  details  are  listed
in Table~1 \cite{flp,flp2,flp3,sll,uemura2,kiefl,kobayashi,murata,taniguchi,bando,tanaka,urayama,rogge,rotter}.
Note that the parameter values vary over several  orders  of
magnitude, which  is  important  for  successful  determination  of  scaling
properties.  We find that $\rho_{\rm s}$ and $T_{\rm c}$ are related to $\sigma_0(T_{\rm c})$ by power  laws  of
the form $\sigma_0^m$  with $m = -1.05(3)$ 
for  $T_{\rm c}$ and $m = -0.77(3)$  for  $\rho_{\rm s}$;  in  both
cases there is a decrease in the strength of  the  superconducting  property
with increasing conductivity.  
For comparison, Fig.~1(b) shows a  similar  plot using the data on cuprates and the elements Nb and Pb recently  reported  by
Homes {\sl et al} \cite{homes}.
Here the overall trend for $T_{\rm c}$ is much less clear and  the trend  for
$\rho_{\rm s}$  shows  a  broad  increase,  opposite to that of the molecular
superconductors, with the positive exponent $m\sim 0.75$.   
This  difference in the $\rho_{\rm s}$--$\sigma_{\rm 0}$ 
scaling between  cuprates  and  molecular  systems  in  the  high
conductivity direction contrasts with the reported similarity in scaling  behaviour
between cuprates  and  organics  in  the   low   conductivity   interplane
direction \cite{dordevic}, corresponding to $m = 0.85$ on a 
$\rho_{\rm s}$--$\sigma_0$ plot.  This value  is
similar to the $m \sim 0.75$ seen for the cuprate high conductivity direction in Fig.~1(b). 
We thus have a  situation  in  which the high and low conductivity directions  in  cuprates,  
along with the low conductivity direction in  layered  organics,  
all share a similar scaling property where $\rho_{\rm s}$ increases with increasing $\sigma_{\rm 0}$, 
whereas for the high conductivity direction in the molecular superconductors $\rho_{\rm s}$ behaves  quite
differently in decreasing with increasing $\sigma_0$.

\begin{figure}[h]
\includegraphics[width=8.3cm]{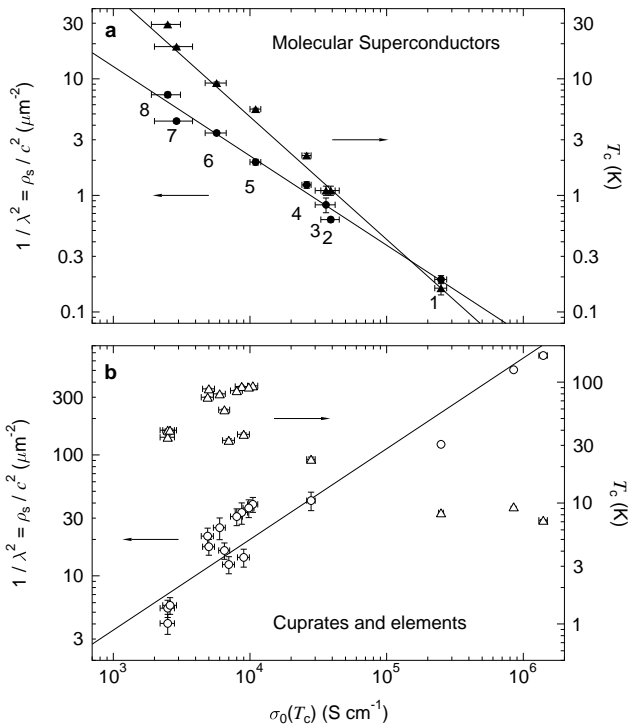}
\caption{
(a) The  inverse  square  of  the  penetration  depth  $1/\lambda^2$(filled
circles, left hand scale)  and  $T_{\rm c}$  
(filled  triangles,  right  hand  scale)
plotted against $\sigma_0(T_{\rm c})$ the normal state conductivity just above  $T_{\rm c}$  in  the
most highly conducting direction.  The key to the molecular  superconductors
is listed in Table~1.  (b)  The  data  of  Homes  {\sl et al} \cite{homes}  
on  cuprates  and
elements for comparison; $1/\lambda^2$(open circles, left hand scale) and  
$T_{\rm c}$  (open
triangles, right hand scale).  From this data  it  can  
be  seen  that  $1/\lambda^2$
exhibits a power law dependence on $\sigma_0$ that is completely  
opposite  to  that
of the molecular systems.
}
\label{fig1}
\end{figure}

\begin{table*}[htbp]
\begin{tabular}{cllll} \hline
Label  &  Material    &  $T_{\rm c}$  & $\lambda$  &  $\sigma_0(T_{\rm c})$ 
\\ 
      &               &      (K)    &  ($\mu$m)    &     ($10^3$ S\,cm$^{-1}$)
\\ \hline
1     &   $\kappa$-BETS$_2$GaCl$_4$   &   0.16(2) \cite{flp}  &  2.3(1) \cite{flp} & 250(25) \cite{kobayashi}
\\
2     &   (TMTSF)$_2$ClO$_4$  &  1.1(1) \cite{flp2} &  1.27(3) \cite{flp2}  &  39(6) (a-axis) \cite{murata}
\\
3     &   $\alpha$-ET$_2$NH$_4$Hg(SCN)$_4$ & 1.1(1) \cite{flp3}   &  1.1(1) \cite{flp3} & 36(6) \cite{taniguchi}
\\
4     &   $\beta$-ET$_2$IBr$_2$  &          2.2(1) \cite{flp3} & 0.90(3) \cite{flp3}  &  26(2) \cite{bando}
\\
5     &   $\lambda$-BETS$_2$GaCl$_4$ &   5.5(1) \cite{flp} & 0.72(2) \cite{flp} &   11(1) \cite{tanaka}
\\
6     &   $\kappa$-ET$_2$Cu(NCS)$_2$ &  9.2(2) \cite{flp3,sll} & 0.54(2) \cite{flp3,sll} & 6(1) \cite{urayama}
\\
7     &   K$_3$C$_{60}$      &      18.9(1) \cite{uemura2} &  0.48(2) \cite{uemura2} &  2.9(9) \cite{rogge}
\\
8     &   Rb$_3$C$_{60}$    &     29.3(1) \cite{kiefl} &  0.42(2) \cite{kiefl} &  2.5(6) \cite{rotter}
\\ \hline
\end{tabular}
\caption{Parameter values for the molecular superconductors.
Values for $T_{\rm c}$ and $\lambda$ are  derived  simultaneously  from  muon  spin  rotation
studies in the vortex state. 
$\lambda$ corresponds to the estimated zero temperature value $\lambda(0)$. 
$\sigma_0(T_{\rm c})$ is  the  normal  state  conductivity  
in  the  most  highly conducting direction. 
The  conductivity  is  derived  from  reported  multi-contact 
resistance measurements in the case of the  organics,  single-domain
STM  measurements  for  K$_3$C$_{60}$  
and  far-infrared  reflectivity  for  Rb$_3$C$_{60}$.
Estimated  uncertainties  in  the  least  significant  digit  are  shown  in
brackets after each value.}
\end{table*}

In the case of the molecular systems, the different power laws seen  for  $\rho_{\rm s}$
and $T_{\rm c}$  against $\sigma_{\rm 0}$ in Fig.~1(a) imply that the scaling between them will not be  of  the
linear Uemura form\cite{uemura} but will follow another power law.  
Fig.~2 shows the Uemura plot of $T_{\rm c}$ against $1/\lambda^2$ 
in log--log form  where  it  can
be seen that $T_{\rm c}$ follows $\rho_{\rm s}^m$ with the  fitted  value  $m  =  1.44(3)$.   
This approximate scaling of $T_{\rm c}$ with $\rho_{\rm s}^{3/2}$, 
or equivalently  $\lambda^{-3}$,  in  2D  organic
superconductors was noted previously  and  discussed  in  terms  of  the  2D
physics of layered superconductors \cite{flp,flp2,powell}.
However, it now appears that  the
scaling  relations  between  $T_{\rm c}$,  $\rho_{\rm s}$  and  
$\sigma_{\rm 0}$   are   more   universal,
encompassing examples of q1D and 3D molecular superconductors alongside  the
2D systems.  The non-linear scaling between $T_{\rm c}$ and 
$\rho_{\rm s}$ in the molecular  case
is much harder to understand than the linear scaling seen in  the  cuprates.
In the cuprates the carrier density $n$ is directly controlled by  the  doping
level; in the underdoped regime $\rho_{\rm s}$ is directly 
proportional to $n$ and $T_{\rm c}$ has been
suggested to be linked to $\rho_{\rm s}$ either through Bose-Einstein condensation of preformed pairs \cite{uemura-bec} or through a mechanism in which phase fluctuations of the superconducting order parameter determine $T_{\rm c}$ \cite{emery2}.
In contrast,  for  the
molecular systems $n$ is fixed by the unit cell size and stoichiometry of  the
crystal structure and varies only little  across  the  range  of  materials,
whose superconducting parameters are  nevertheless  varying  across  several
orders of magnitude.  Differences in  the  superconducting  properties  must
then arise entirely from differences in the details of the electronic  many--body interactions.

\begin{figure}[h]
\includegraphics[width=8.3cm]{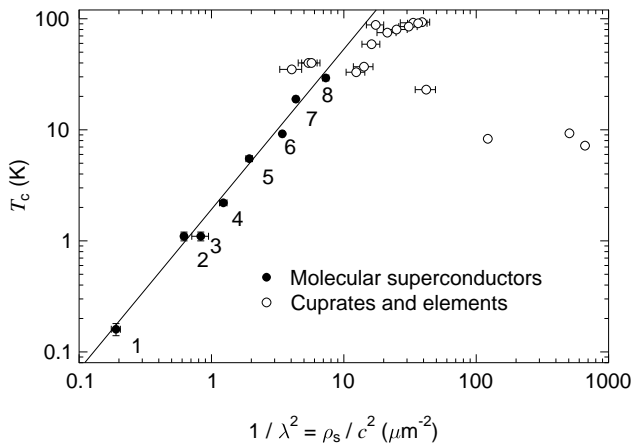}
\caption{
Log-log Uemura plot of $T_{\rm c}$ against $1/\lambda^2$. 
Data for the  cuprates  and
elemental superconductors tabulated by Homes  {\sl et al} \cite{homes}  
are  also  shown  for
comparison. For the molecular systems a scaling close to 
$\rho_{\rm s}^{3/2}$ is  observed,
rather than the linear $\rho_{\rm s}$  scaling seen for the cuprates.
}
\label{fig2}
\end{figure}

Further evidence for fundamentally  different  behaviour  between  molecular
and non-molecular superconductors is seen when an attempt is made to  search
for linear scaling between $\rho_{\rm s}$ and the product 
$\sigma_0(T_{\rm c})T_{\rm c}$,  of  the  form  that
was recently demonstrated by Homes {\sl et al} \cite{homes}.  
Fig.~3(a) shows that such a simple linear
scaling does not occur  for  the  molecular  superconductors.     The
linear behaviour seen for the non-molecular systems can be  understood  from
applying the Ferrell-Glover-Tinkham sum  rule  for  the  real  part  of  the
frequency dependent conductivity \cite{ferrell},
\begin{equation}
{2 \over \pi} \int_0^\infty \sigma(\omega)\,{\rm d}\omega = {ne^2 \over
  m^*}
= \epsilon_0 \rho_{\rm s},
\end{equation}
 where  $\sigma(\omega)$ takes the Drude form $\sigma_0 /  (1+
(\omega/\Gamma)^2)$, with $\sigma_0 = n e^2 / (m^* \Gamma)$ and $\Gamma$ being the scattering rate.  
In the case where $\Gamma$  is  significantly
smaller than the frequency corresponding to the superconducting  energy  gap
$2\Delta/\hbar$, the whole free carrier spectrum is redistributed to zero  frequency  to  give  
the  superfluid response peak, i.e. $\epsilon_0\rho_{\rm s} = \sigma_0(T_{\rm c})\Gamma$.  
If, on the other  hand,  $2\Delta/\hbar$  is  significantly
smaller than $\Gamma$,  
then  the  normal  state  conductivity  is  independent  of
frequency in the gap region, i.e.\ $\sigma(\omega,T_{\rm
  c})=\sigma_0(T_{\rm c})$; in this case, as  the  superconducting
gap forms, an area of the conductivity spectrum with  frequency  width 
$2\Delta/\hbar$ 
and height $\sigma_0$ is redistributed to zero frequency to give  
the  superfluid response peak. 
This leads to the following expression for 
$\epsilon_0\rho_{\rm s}$:
\begin{equation}
\epsilon_0\rho_{\rm s} = {2 \over \pi} \sigma_0(T_{\rm c}) {2\Delta \over
  \hbar}
= { 2 k_{\rm B} \over \pi\hbar } \eta\, \sigma_0(T_{\rm c}) T_{\rm c}, 
\label{eq}
\end{equation}
where
$\eta = 2\Delta/k_{\rm B}T_{\rm c}$.  
In Fig.~3(a) the dashed line shows Eqn.~\ref{eq} plotted  taking
the weak-coupling BCS limit $\eta =  3.53$  as  a  reference;  
this  is  seen  to describe the general behaviour of the non-molecular  data  quite  well.  
The effective value of $\eta$ derived from the data using Eqn.~\ref{eq} is shown  in  Fig.~3(b),
which reveals the considerable variation among the molecular systems.   
Note that Eqn.~\ref{eq} was derived  assuming that  the  ratio  of  carrier  density  to
effective mass is the same in the normal  and  superconducting  states.  If,
however, this assumption is relaxed then the effective  gap  ratio  observed
in this plot becomes
\begin{equation}
\eta = \left( {2\Delta\over k_{\rm B}T_{\rm c}} \right)
\left( { n_{\rm s} \over n_{\rm n} } \right)
\left( { m^*_{\rm n} \over m^*_{\rm s} } \right),
\label{eta}
\end{equation}
where the subscripts $s$ and $n$ refer to the superconducting and normal  states
respectively.  
Strong coupling can increase $\eta$ over the  BCS  value  via  the first term of Eqn.~\ref{eta}, 
however if Eqn.\ref{eq} is applicable to the molecular systems then the reduced values of $\eta$ seen
for many cases would require a  contribution  from  at  least
one of the other two terms, i.e. the superconducting carrier  density  would
need to be less than normal state carrier density or the effective  mass  of
superconducting carriers would have to be larger than that of  the  carriers
in the normal state.

\begin{figure}[h]
\includegraphics[width=8.3cm]{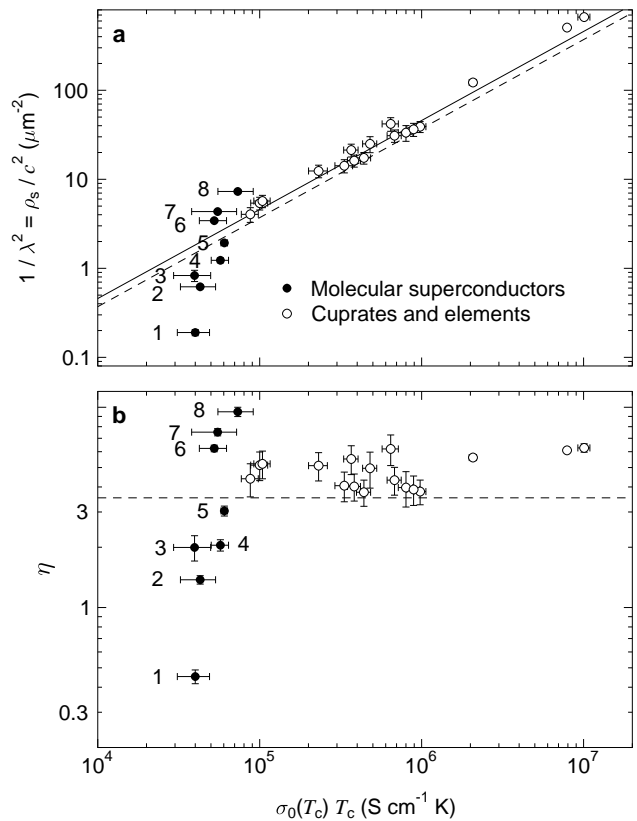}
\caption{
(a) Plot of $1/\lambda^2$ against the product of $T_{\rm c}$  
and  $\sigma_0(T_{\rm c})$,  following
Homes {\sl et al} \cite{homes}.  
For the molecular superconductors the data  collapse  onto  a
narrow ordinate  range  due  to  the  inverse  scaling  between  
$T_{\rm c}$  and  $\sigma_0$
demonstrated in Fig.~1(a). 
Open circles show the data of  Homes  {\sl et al} \cite{homes}  along
with the linear  fit  (solid  line).  The  dashed  line  shows  the  scaling
expected for a weak coupling BCS superconductor in the high scattering rate limit (Eqn.\ref{eq}). 
(b) The data  expressed  as
an effective gap parameter $\eta = (2\Delta/k_{\rm B}T_{\rm c}) 
(n_{\rm s} / n_{\rm n}) (m_{\rm n}^* /  m_{\rm s}^*)$.   Whereas
the cuprates and elements are grouped around a value of  $\eta$
just  above  the
BCS limit (dashed line) and comparable to the gap ratios  seen  using  other
techniques, the molecular systems cover a wide  range  of $\eta$  
values,  both
above and below the BCS limit.
}
\label{fig3}
\end{figure}

Another way to look at the data is to plot the ratio $\epsilon_0\rho_{\rm s}/\sigma_0$ which  gives  a
measure  of  the  effective  frequency  width  $\Gamma_{\rm e}$  of   the   normal   state
conductivity spectrum that provides the superfluid  response  (Fig.~4).
This will be determined either by $\Gamma$ itself  or by  $(2/\pi)(2\Delta/\hbar)$, whichever is the smaller.  
For the non-molecular systems  the  effective width  follows the linear $T_{\rm c}$ dependence expected if it is 
proportional to  either  a  BCS-type gap or a $T$-linear scattering rate. 
In contrast, for  the  molecular  systems $\Gamma_{\rm e}$ follows a steeper power law $T_{\rm c}^\alpha$ 
with the fitted value $\alpha = 1.58(5)$.   
This behaviour suggests that the molecular systems are in the low-scattering-rate limit where 
$\Gamma  <  (2/\pi)  2\Delta/\hbar$ and $\Gamma_{\rm e}$  follows $\Gamma$.
We note that the fitted power law for  $\Gamma_{\rm e}$  is  also  broadly  consistent
with the temperature dependence of the  scattering  rate  deduced  from  the
temperature dependent  resistance  of  individual  examples  of  the
molecular  metals; measurements for molecular metals just above T$_{\rm c}$ generally  show   power-law
exponents in the region 1.5 to 2 \cite{kobayashi,murata,taniguchi,bando,tanaka,urayama,rogge}. 

\begin{figure}[h]
\includegraphics[width=8.3cm]{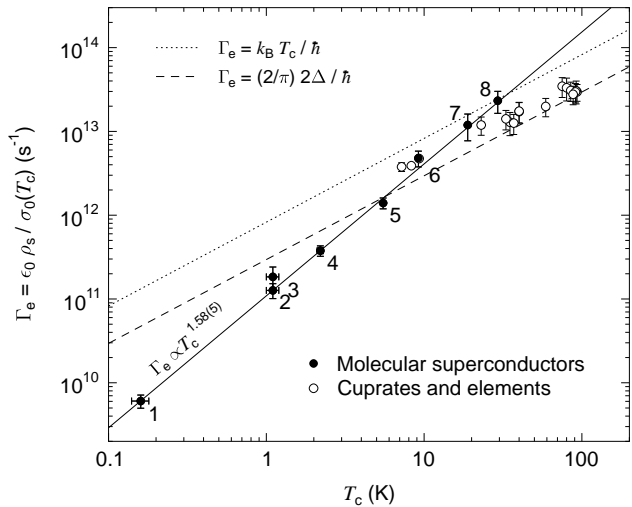}
\caption{
Plot of $\rho_{\rm s}$ normalised by 
$\sigma_0(T_{\rm c})/\epsilon_0$ to give the frequency $\Gamma_{\rm e}$.  
This
reflects the effective width of  the  normal  state  conductivity  spectrum,
$\sigma(\omega)$, 
that has condensed into the superfluid peak at $\omega=0$. 
For  the  cuprates
and elements, $\Gamma_{\rm e}$ appears to follow  a  
linear  temperature  dependence.   In
contrast,  for  the  molecular  superconductors  the   steeper   temperature
dependence $\Gamma_{\rm e} \propto T_{\rm c}^{1.58(5)}$ 
is seen (solid line).  $\Gamma_{\rm e}$  for  the  cuprates  and
elements and higher $T_{\rm c}$ molecular superconductors is 
seen to lie between  the
BCS weak-coupling limit (dashed line, Eqn.~\ref{eq}) 
and the  Planckian  scattering
rate limit \cite{zaanen} (dotted line).
}
\label{fig4}
\end{figure}

The scaling relations for the  molecular superconductors  highlighted  here
suggest that there are features of  their  electronic  properties  that  are
common, despite the various materials having quite different  dimensionality
and Fermi surface topology.   
The simplicity of the  scaling  also  suggests
that it is being controlled by a single  dominant  parameter,  such  as  the
ratio of the electron correlation energy on a molecule $U$ to the  electronic
bandwidth $W$. 
$U/W$ also controls the proximity of the Mott insulator (MI) phase;
being close to the MI phase supports higher $T_{\rm c}$
(as seen in studies of the $\kappa$-phase ET salts under pressure \cite{Schirber,Caulfield}) and 
the inverse relation between $T_{\rm c}$ and $\sigma_{\rm 0}$ follows naturally from this.
However standard approaches to modelling the crossover between MI and superconducting phases predict that the enhanced $T_{\rm c}$ near the MI phase is accompanied by a depressed $\rho_{\rm s}$ \cite{benslatest}, 
exactly the opposite of what is observed experimentally.   
Identification of new theoretical models that match the observed
scaling  behaviour is clearly necessary; finding such models should lead to significant progress in understanding 
superconductivity in molecular systems.

We  acknowledge  discussions  and  interaction  with   Ben
Powell, Ross McKenzie, Nikitas Gidopoulos,  Steve  Lee,   Naoki  Toyota  and
Takahiko Sasaki.  We are also grateful to  Katherine  Blundell  for  helpful
comments on the manuscript and to  the  staff  at  the  PSI  and  ISIS  muon
facilities for research support.


\begin{thebibliography}{00}
\bibitem{uemura}
Y.J. Uemura  {\sl et al.},   
%% Universal  correlations  between  $T_{\rm c}$  and  ns/m*
%% (carrier density over effective mass) in high  $T_{\rm c}$  cuprate  superconductors.
Phys. Rev. Lett. {\bf 62}, 2317 (1989).
\bibitem{homes} %2. 
C.C. Homes {\sl et al.}, 
%% A universal scaling relation in high-temperature
%% superconductors. 
Nature {\bf 430}, 539 (2004).
\bibitem{zaanen} %3
J. Zaanen, 
%% Why the temperature is high, 
Nature {\bf 430}, 512 (2004).
\bibitem{emery} %4
V.J. Emery and S.A. Kivelson, 
%Superconductivity in bad metals, 
Phys. Rev. Lett. {\bf 74}, 3253 (1995).
\bibitem{sjb} %5
S.J. Blundell, 
%Spin-polarized muons in condensed matter physics,
Contemp. Phys. {\bf 40}, 175 (1999).
\bibitem{sonier} %6
J.E. Sonier, J.H. Brewer and R.F. Kiefl, 
%?SR studies of the vortex state in type-II superconductors, 
Rev. Mod. Phys. {\bf 72}, 769 (2000).
\bibitem{flp} %7
F.L. Pratt {\sl et al.}, 
%?SR studies of magnetic superconductors based on the BETS molecule, 
Polyhedron {\bf 22}, 2307 (2003).
\bibitem{flp2} %8
F.L. Pratt {\sl et al.}, 
%?SR studies of layered organic superconductors:
%vortex phases, penetration depth and anomalous superfluid properties,
Synth. Met.  (in press).
\bibitem{flp3} %9
F.L. Pratt {\sl et al.}, 
%BEDT-TTF superconductors studied by ?SR, 
Physica B {\bf 289--290}, 396 (2000).
\bibitem{sll} %10
S.L. Lee {\sl et al.}, 
%Investigation of vortex behaviour in the organic
%superconductor ?-(BEDT-TTF)2Cu(SCN)2 using muon spin rotation, 
Phys. Rev.
Lett. {\bf 79}, 1563 (1997).
\bibitem{uemura2} %11
Y.J. Uemura {\sl et al.}, 
%Magnetic-field penetration depth in K3C60 measured by muon spin
%relaxation, 
Nature {\bf 352}, 605 (1991).
\bibitem{kiefl} %12
R.F. Kiefl {\sl et al.}, 
%Coherence peak and superconducting energy gap in
%Rb3C60 observed by muon spin relaxation, 
Phys. Rev. Lett. {\bf 70}, 3987
(1993).
\bibitem{kobayashi} %13
A. Kobayashi, T. Udagawa, H. Tomita, T. Naito and H. Kobayashi,
%New organic metals based on BETS compounds with MX4- anions (BETS =
%bis(ethylenedithio)tetraselenafulvalene; M = Ga, Fe, In; X = Cl, Br), 
Chem.
Lett. {\bf 1993}, 2179 (1993).
\bibitem{murata} %14
K. Murata, H. Anzai, G. Saito, K. Kajimura and T. Ishiguro, 
%Evidence
%for three dimensional ordering of superconductivity in highly anisotropic
%organic conductor, 
%(TMTSF)2ClO4, 
J. Phys. Soc. Jpn. {\bf 50}, 3529 (1981).
\bibitem{taniguchi} %15
H. Taniguchi, Y. Nakazawa and K. Kanoda, 
%Phase diagram of vortices
%in the quasi-two-dimensional organic superconductor ?-(BEDT-
%TTF)2NH4Hg(SCN)4: a system of pancake vortices with out-of-plane coupling
%dominated by the electromagnetic energy, 
Phys. Rev. B {\bf 57}, 3623 (1998).
\bibitem{bando} %16
H. Bando {\sl et al.}, 
%Ordinary and anomalous magnetoresistances in ?-(BEDT-
%TTF)2X (X=IBr2, I2Br, I3), 
J. Phys. Soc. Jpn. {\bf 54}, 4265 (1985).
\bibitem{tanaka} %17
H. Tanaka, A. Kobayashi, A. Sato, H. Akutsu and H. Kobayashi,
%Chemical control of electrical properties and phase diagram of a series of
%?-type BETS superconductors, ?-(BETS)2GaBrxCl4-x, 
J. Am. Chem. Soc. {\bf 121}, 760 (1999).
\bibitem{urayama} %18
H. Urayama {\sl et al.}, 
%Crystal and electronic structures and physical
%properties of $T_{\rm c}$
%=10.4 K superconductor (BEDT-TTF)2Cu(NCS)2, 
Synth. Met. {\bf 27},
A393 (1988).
\bibitem{rogge} %19
S. Rogge, M. Durkut and T.M. Klapwijk,  
%Single domain transport
%measurements of C60 films,  
Phys. Rev.  B {\bf 67}, 033410 (2003).
\bibitem{rotter} %20
L.D. Rotter  {\sl et al.}, 
%Infrared reflectivity measurements of a
%superconducting energy scale in Rb3C60, 
Nature {\bf 355}, 532 (1992).
\bibitem{dordevic} %21
S.V. Dordevic {\sl et al.}, 
%Global trends in the interplane penetration depth
%of layered superconductors, 
Phys. Rev. B {\bf 65}, 134511 (2002).
\bibitem{flp4} %22
F.L. Pratt, S.J. Blundell, T. Lancaster, S.L. Lee and 
N. Toyota,
%Electrodynamics of molecular organic superconductors studied by ?SR, 
J. Phys. IV France {\bf 114}, 367 (2004).
\bibitem{powell} %23
B.J. Powell and R.H. McKenzie, 
%On the relationship between the
%critical temperature and the London penetration depth in layered organic
%superconductors, 
J. Phys.: Condens. Matter {\bf 16}, L367 (2004).
\bibitem{uemura-bec}
Y.J. Uemura, Physica {\bf C282-287}, 194 (1997).
\bibitem{emery2} %24
V.J. Emery  and S.A. Kivelson, 
%Importance of phase fluctuations in
%superconductors with small superfluid density, 
Nature {\bf 374}, 434 (1995).
\bibitem{ferrell} %25
R.A. Ferrell and R.E. Glover, 
%Conductivity of superconducting films:
%a sum rule, 
Phys. Rev. {\bf 109}, 1398 (1958).
\bibitem{Schirber}
J.E. Schirber {\sl et al}, Phys. Rev. B {\bf 44}, 4666 (1991).
\bibitem{Caulfield}
J. Caulfield {\sl et al}, J. Phys.:Condens. Matter {\bf 6}, 2911 (1994).
\bibitem{benslatest}
B.J. Powell and R.H. McKenzie, cond-mat/0410125.


\end{thebibliography}
\end{document}